\documentstyle[psfig]{mn}

\newif\ifAMStwofonts
\AMStwofontstrue

\def\spose#1{\hbox to 0pt{#1\hss}}
\def \h121cm{H{\sc \,i}\,21\,cm}
\def \dy{\Delta y/y}
\def \da{\Delta\alpha/\alpha}
\def \txs{TXS\,0218$+$357}
\def \pks{PKS\,1413$+$135}
\def \b3{B3\,1504$+$377}

\title[Improved constraints on possible variation of physical constants
from QSO absorption lines]{Improved constraints on possible variation of
physical constants from \h121cm and molecular QSO absorption lines}

\author[M. T. Murphy et al.]{M. T. Murphy$^1$\thanks{E-mail:
       mim@phys.unsw.edu.au (MTM)}, J. K. Webb$^1$, V. V. Flambaum$^1$,
       M. J. Drinkwater$^2$,\newauthor F. Combes$^3$ and T. Wiklind$^4$\\
$^1$School of Physics, The University of New South Wales, UNSW Sydney NSW 2052, Australia\\
$^2$School of Physics, University of Melbourne, Victoria 3010, Australia\\
$^3$DEMIRM, Observatoire de Paris, 61 Av. de l'Observatoire, Paris, F-75014, France\\
$^4$Onsala Space Observatory, Onsala, S-43992, Sweden}

\date{Accepted ---.
      Received ---;
      in original form ---}

\pagerange{\pageref{firstpage}--\pageref{lastpage}}
\pubyear{2001}

\begin{document}

\maketitle

\label{firstpage}

\begin{abstract}
QSO absorption spectra provide an extremely useful probe of possible
cosmological variation in various physical constants. Comparison of \h121cm
absorption with corresponding molecular (rotational) absorption spectra
allows us to constrain variation in $y\equiv\alpha^2g_p$ where $\alpha$ is
the fine structure constant and $g_p$ is the proton $g$-factor. We analyse
spectra of two QSOs, \pks~and \txs, and derive values of $\dy$ at
absorption redshifts of $z=0.2467$ and $0.6847$ by simultaneous fitting of
the \h121cm and molecular lines. We find $\dy=(-0.20\pm 0.44)\times
10^{-5}$ and $\dy=(-0.16\pm 0.54)\times 10^{-5}$ respectively, indicating
an insignificantly smaller $y$ in the past. We compare our results with
other recent constraints from the same two QSOs (Drinkwater et al. 1998;
Carilli et al. 2000) and with our recent optical constraints which
indicated a smaller $\alpha$ at higher redshifts.
\end{abstract}


\section{Introduction}\label{sec:intro}
Currently, many modern unified theories provide strong motivation for
experimental searches for variation in physical constants. Kaluza-Klein,
Superstring and M-theory require the existence of extra spatial dimensions
which are compactified on small scales. The coupling constants in our
3-dimensional subspace of these theories are related to the scale sizes of
the extra dimensions. Therefore, if the extra dimensions evolve with
cosmology, we expect our 3-dimensional constants to vary (e.g. Forg\'{a}cs
\& Horv\'{a}th 1979; Marciano 1984; Barrow 1987; Damour \& Polyakov
1994). Currently, there seems to be no mechanism for keeping the sizes of
the extra dimensions constant (Li \& Gott 1998) and so these modern
theories naturally predict variations in the fundamental constants.

Experimentally, quasar (QSO) absorption lines provide an ideal probe of
cosmological variation of physical constants. Savedoff (1956) first
analysed doublet separations seen in galaxy emission spectra to obtain
constraints on variation in the fine structure constant, $\alpha \equiv
e^2/\hbar c$. Absorption lines in intervening clouds along the line of
sight to QSOs are substantially narrower than intrinsic emission lines and
so provide a much more precise probe of $\alpha$ at high redshift. Bahcall,
Sargent \& Schmidt (1967) first used the doublet spacings of gas seen in
absorption against background QSOs to derive firm upper limits on possible
variation of $\alpha$. We summarize results from optical absorption line
studies in Section \ref{sec:discussion}, suffice it to say here that the
tightest constraint on $\da$ found to date is a tentative detection of
variation at $\da = (-0.72 \pm 0.18) \times 10^{-5}$ in 49 absorption
clouds over the redshift range $0.5 < z < 3.5$ (Murphy et al. 2001a; Webb
et al. 2001).\footnote{See Murphy et al. (2001a) for a summary of all
relevant constraints on $\da$.}

Even tighter constraints can be placed on variability of combinations of
physical constants using radio QSO absorption spectra. A comparison of
\h121cm and molecular rotational absorption spectra is particularly
convenient. The \h121cm hyperfine transition frequency is proportional to
$\mu_p\mu_B/(\hbar a^3)$ where $\mu_p=g_pe\hbar/(4m_pc)$ and
$\mu_B=e\hbar/(2m_ec)$ is the Bohr magneton. Here, $g_p$ is the proton
g-factor, $m_p$ and $m_e$ are the masses of the proton and electron
respectively and $a=\hbar^2/(m_ee^2)$ is the Bohr radius. The rotational
transition frequencies of diatomic molecules, such as CO, are proportional
to $\hbar/(Ma^2)$ where $M$ is the reduced mass. Therefore, the ratio of
the hyperfine and molecular rotational frequencies is proportional to
$\alpha^2g_pM/m_p$. Variations in this quantity will be dominated by
variations in $y\equiv\alpha^2g_p$ since variations in $M/m_p$ are
suppressed by a factor $m_p/U \sim 100$ where $U$ is the binding energy of
nucleons in nuclei. If any variation in $y$ occurs, it will be observed as
a difference in redshift between the \h121cm ($z_{\rm H}$), and molecular
($z_{\rm mol}$) absorption lines:
\begin{equation}\label{eq:dy}
\frac{\Delta y}{y}\equiv \frac{y_z-y_0}{y_0} \approx \frac{\Delta z}{1+z}
\equiv \frac{z_{\rm mol}-z_{\rm H}}{1+z_{\rm mol}}
\end{equation}
where $y_z$ and $y_0$ are the values of $y$ at the absorption redshift $z$
and in the laboratory respectively.

The first comparison of \h121cm and molecular absorption was made by
Varshalovich \& Potekhin (1996). They compared the published redshifts of
the CO absorption (reported in Wiklind \& Combes 1994) and \h121cm
absorption (reported in Carilli, Perlman \& Stocke 1992) towards \pks. They
interpreted any shift as a change in the molecular mass, $M$. Wiklind \&
Combes (1997) carried out a similar analysis. Drinkwater et al. (1998,
hereafter D98) pointed out that such a shift actually constrains $\dy$
rather than $\Delta M/M$ with a similar argument to that
above. Varshalovich \& Potekhin's (1996) value of $\Delta M/M$ translates
to $\dy = (-4 \pm 6) \times 10^{-5}$.

Like Varshalovich \& Potekhin (1996), in D98 we used the published
redshifts for the CO(0$-$1) lines towards \pks\footnote{The CO data of
Wiklind \& Combes (1994), used by Varshalovich \& Potekhin (1996), showed a
$-11{\rm \,kms}^{-1}$ offset from the \h121cm velocity. We used a new,
corrected CO data set reported in Wiklind \& Combes (1997) which shows no
such offset.}. However, we improved the precision of $\dy$ by fitting Voigt
profiles to the \h121cm data rather than using the published redshifts. The
errors on the redshift of each velocity component in the H{\sc \,i} data
were reduced by an order of magnitude compared to the estimates of
Varshalovich \& Potekhin (1996). We also applied this analysis to another
absorber towards \txs~at $z_{\rm abs}=0.685$. In D98 we derived an upper
limit on any variation in $y$ for both absorbers: $\left|\dy\right| <
0.5\times 10^{-5}$.

The present work aims at turning our upper limit on $\dy$ in D98 into a
measurement of $\dy$ by fitting simultaneously the \h121cm data {\it and}
the rotational lines of several molecular species. We describe the
available data in Section 2 and describe our analysis and results in
Section 3. We discuss our new results in Section 4, comparing them with
other constraints on $\alpha$ variability.

\section{Available data}\label{sec:data}
Only 4 QSOs have had mm-band rotational molecular absorption detected along
their lines of site: \txs, \pks, \b3 and PKS\,1830$-$211. Below we describe
the molecular and H{\sc \,i} data available for the first two of these
QSOs. We do not have sufficient data to consider PKS\,1830$-$211. The
molecular absorption profile is very broad (${\rm FWHM}\sim 40{\rm
\,kms}^{-1}$, Wiklind \& Combes 1998) and so constraints on $\dy$ would be
weak. We do not consider \b3 any further due to a $15{\rm \,kms}^{-1}$
offset between the \h121cm (Carilli et al. 1997, 1998) and HCO$^+$ (Wiklind
\& Combes 1996) velocity scales as noted in D98. Checks on the origin of
this difference have not revealed any instrumental or human error and so it
may well be that the molecular and H{\sc \,i} absorption occurs in
different clouds. However, as noted in D98, this seems unlikely because the
molecular and H{\sc \,i} velocity structures are very similar. Further data
is required to clarify this problem.

\subsection{\txs}\label{sec:txs}
\txs~is a gravitationally lensed QSO (Patnaik et al. 1993), probably at a
redshift $z_{\rm em}\approx 0.94$ (Wiklind \& Combes 1995), showing
absorption in the lensing galaxy at $z_{\rm abs} = 0.6847$ (Carilli, Rupen
\& Yanny 1993). VLBI observations (Patnaik et al. 1993; Biggs, Browne \&
Wilkinson 2001) show two, compact, flat-spectrum components (A to the SW
and B to the NE) and a steep-spectrum Einstein ring. The background QSO
shows intensity variability on intraday (Biggs et al. 2001) and longer
($\sim$monthly) time-scales (O'Dea et al. 1992; Patnaik et al. 1993). We
use the spectrum of this QSO despite the fact that it is lensed since the A
component provides the dominant absorption in both the molecular (Menten \&
Reid 1996) and 21\,cm (Carilli et al. 2000, hereafter C00) bands. Thus, the
profile of the $z=0.6847$ absorber should not vary significantly over the
time between molecular and \h121cm~observations.

In this paper we use the \h121cm spectrum of the $z=0.6847$ absorption
system published by Carilli et al. (1993) (${\rm FWHM}=6.9{\rm
\,kms}^{-1}$). Spectra of CO(1$-$2), $^{13}$CO(1$-$2), C$^{18}$O(1$-$2) and
CO(2$-$3) are taken from Combes \& Wiklind (1995) (${\rm FWHM}\sim{\rm
30\,ms}^{-1}$) and those of HCO$^+$(1$-$2) and HCN(1$-$2) are from Wiklind
\& Combes (1995) (${\rm FWHM}\sim{\rm 2\,kms}^{-1}$). These data are
presented in Fig. 1.

\begin{figure*}
\centerline{\psfig{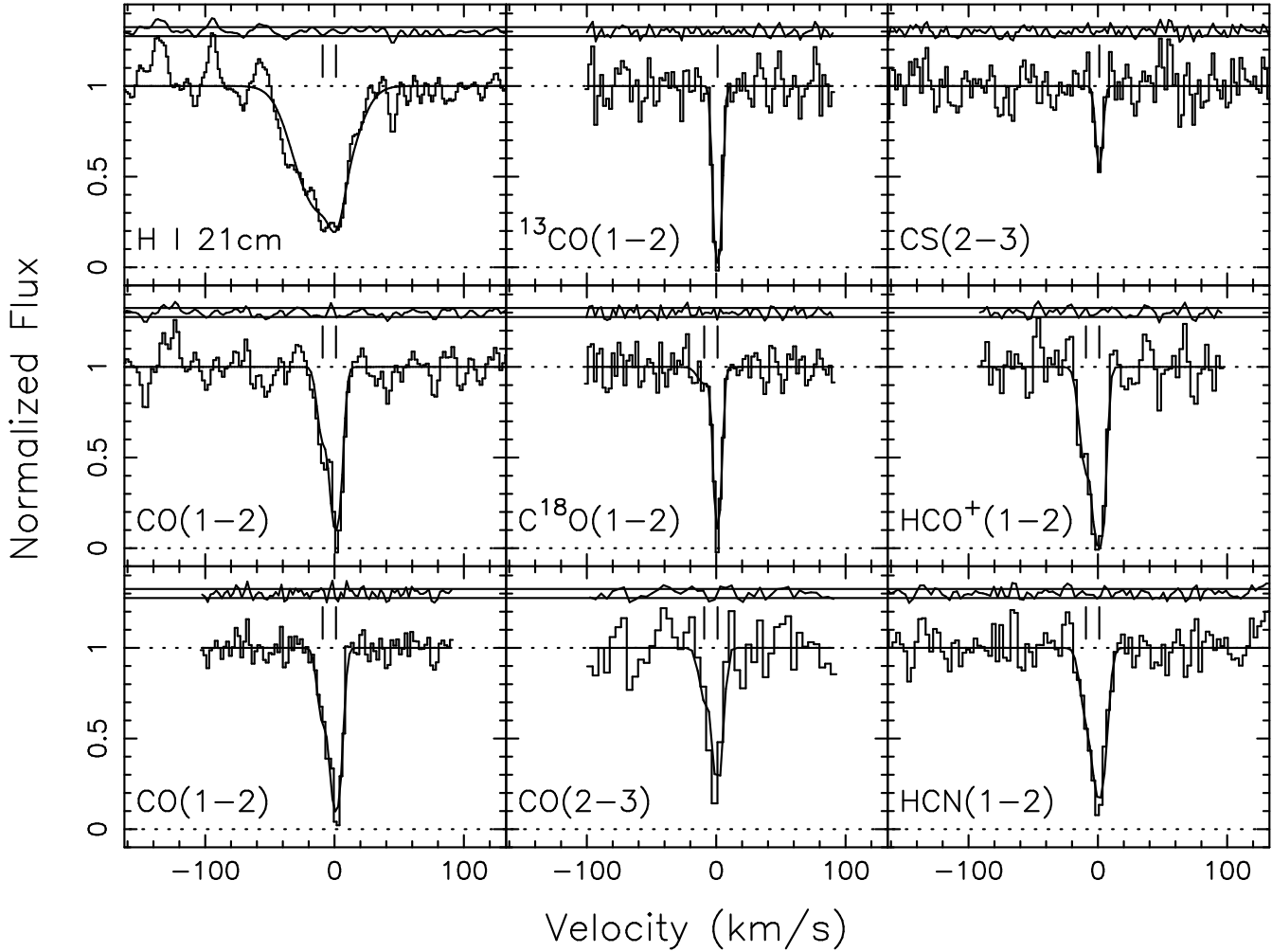}}
\caption{Spectra of the $z=0.6847$ absorption system towards \txs. The data
are plotted as a histogram and the solid line represents our Voigt profile
fit. The tick-marks above the data show the position of the fitted velocity
components. The residuals (i.e. [data]-[fit]), normalized to a constant
1$\sigma$-error (defined by the signal-to-noise ratio in the continuum),
are shown above each spectrum. Note that two independent spectra of
CO(1$-$2) have been included.}
\end{figure*}

\subsection{\pks}\label{sec:pks}
\pks~is at the centre of an edge on spiral galaxy at $z_{\rm em}=0.24671$
and the absorption occurs in the disk of this galaxy (Wiklind \& Combes
1997). VLBI observations (C00) show that the 21\,cm continuum flux is
dominated by a jet extending from an inverted-spectrum nucleus. The nucleus
dominates the mm continuum emission (Perlman et al. 1996). Thus, the
\h121cm and molecular absorption lie along different sight lines separated
by $\sim 0\farcs03$ (C00). We address this problem and its effect on our
results in Section \ref{sec:discussion}.

Here we use the \h121cm spectrum of the $z=0.2467$ absorption system
observed by Carilli et al. (1992) (${\rm FWHM}\approx{\rm
1.3\,kms}^{-1}$). The molecular absorption lines of CO(0$-$1),
HCO$^+$(1$-$2) and HCO$^+$(2$-$3) (${\rm FWHM}\sim{\rm 40\,ms}^{-1}$) were
observed with the narrow band autocorrelator at the IRAM 30\,m telescope on
Pico Veleta. The details of these observations are reported in Wiklind \&
Combes (1997). These data are presented in Fig. 2.

\begin{figure*}
\centerline{\psfig{file=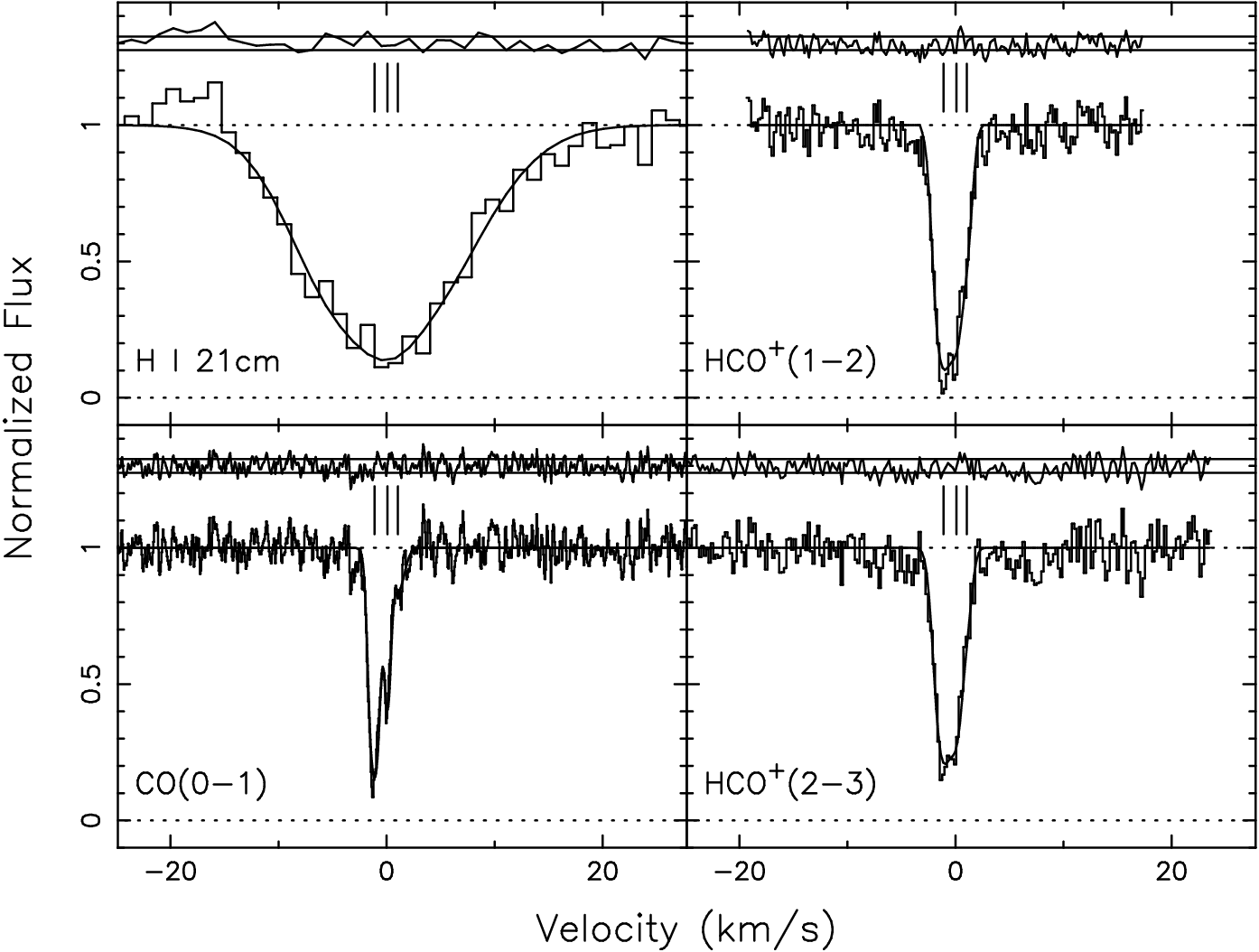,width=7in,angle=270}}
\caption{Spectra of the $z=0.2467$ absorption system towards \pks. The data
are plotted as a histogram and the solid line represents our Voigt profile
fit. The tick-marks above the data show the position of the fitted velocity
components. The residuals (i.e. [data]-[fit]), normalized to a constant
1$\sigma$-error (defined by the signal-to-noise ratio in the continuum),
are shown above each spectrum.}
\end{figure*}

\section{Analysis, results and errors}\label{sec:results}
We used {\sc vpfit}\footnote{See
http://www.ast.cam.ac.uk/$^{\sim}$rfc/vpfit.html for details about
obtaining {\sc vpfit}.} to fit multi-velocity component Voigt profiles to
both the \h121cm~and molecular absorption lines. The laboratory values for
the transition frequencies were taken from Pickett et
al. (1998)\footnote{See http://spec.jpl.nasa.gov} and Essen et al. (1971)
and are given in Table \ref{tab:freqs}. By simultaneously fitting several
molecular transitions of differing strengths in the same absorber, we get
the best possible estimate of the velocity structure.  We determine the
number of velocity components to be fitted to each system by requiring that
the reduced $\chi^2$ for the fit be $\sim 1$.

We use an iterative technique to find the best fit value of $\dy$. We vary
$\dy$ outside {\sc vpfit} (i.e. we vary the ratio of the hyperfine and
molecular frequencies) and fit all lines simultaneously. During the fit we
force the redshift parameters for corresponding velocity components to be
the same for all transitions, thereby reducing the number of free
parameters. We change $\dy$ at each iteration of the routine, $i$, to find
$\chi^2_i$ as a function of $\dy$. The best fit value of $\dy$ is that
which gives the minimum $\chi^2$, $\chi^2_{\rm min}$, and the 1$\sigma$
error in $\dy$ is found using $\chi^2_{\rm min}+1$.

\begin{table}
\begin{center}
\caption{Laboratory frequencies of transitions used in our
analysis. Molecular frequencies are taken from Pickett et al. (1998) and
the \h121cm frequency is from Essen et al. (1971).}
\label{tab:freqs}
\begin{tabular}{lcl}\hline
\multicolumn{1}{c}{Molecule}&\multicolumn{1}{c}{Transition}&\multicolumn{1
}{c}{Frequency/GHz}\\\hline
CO        & 0$-$1 & 115.2712018(5)  \\
          & 1$-$2 & 230.5380000(5)  \\
          & 2$-$3 & 345.7959899(5)  \\
$^{13}$CO & 1$-$2 & 220.3986765(53) \\
C$^{18}$O & 1$-$2 & 219.5603568(81) \\
CS        & 2$-$3 & 146.969033(50)  \\
HCO$^+$   & 1$-$2 & 178.375065(50)  \\
          & 2$-$3 & 267.557619(10)  \\
HCN       & 1$-$2 & 177.261110(2)   \\
H{\sc \,i}& 21\,cm& 1.420405751766(1)\\\hline
\end{tabular}
\end{center}
\end{table}

For the $z=0.6847$ absorption system towards \txs~we find the best fit
value of $\dy=(-0.16\pm 0.36)\times 10^{-5}$ and for the $z=0.2467$
absorption system towards \pks~we find $\dy=(-0.20\pm 0.20)\times
10^{-5}$. To check internal consistency, we have investigated the effect of
removing individual transitions from the analysis of each absorption
system. Our values of $\dy$ showed no significant change upon removal of
any molecular transition. In particular, we found no change when we removed
either or both the CS(2$-$3) and $^{13}$CO(0$-$1) transitions from our
analysis of \txs. The low velocity component does not seem to be present in
these transitions and was not fitted (see Fig. 1). The fact that the
results are insensitive to removal of these lines indicates the robustness
of our line fitting method and highlights the advantage of using many
different molecular transitions.

The 1$\sigma$ errors we quote above are statistical only. However, as noted
in Section \ref{sec:pks}, the fact that we probe slightly different sight
lines with the molecular and \h121cm~observations may result in an
additional random error. D98 have investigated this problem empirically by
fitting several QSO spectra with Galactic absorption and comparing the
fitted redshifts of the HCO$^+$ and H{\sc \,i} velocity components. D98
find a close correspondence between the velocity components of HCO$^+$ and
H{\sc \,i} with a Gaussian dispersion of only $1.2{\rm \,kms}^{-1}$. This
corresponds to an error $\dy=0.4\times 10^{-5}$. Our two results, being of
similar value and so close to zero, suggest that this error estimate may be
too large. However, it is clear that such a conclusion can only be reached
with a larger sample of absorption systems and so we have added this error
in quadrature to obtain our final results:
\begin{equation}\label{eq:resy}
\begin{array}{l}
\hspace{-1.7mm}\dy = (-0.20\pm 0.44)\times 10^{-5}{\rm ~~at~}z=0.2467{\rm~~and}\\
\hspace{-1.7mm}\dy = (-0.16\pm 0.54)\times 10^{-5}{\rm ~~at~}z=0.6847\\
\end{array}
\end{equation}
or
\begin{equation}\label{eq:resa}
\begin{array}{l}
\hspace{-1.7mm}\da = (-0.10\pm 0.22)\times 10^{-5}{\rm ~~at~}z=0.2467{\rm~~and}\\
\hspace{-1.7mm}\da = (-0.08\pm 0.27)\times 10^{-5}{\rm ~~at~}z=0.6847\\
\end{array}
\end{equation}
assuming (perhaps incorrectly) a constant $g_p$.

\section{Discussion}\label{sec:discussion}
Very recently, C00 has reported upper limits on $\dy$ using the same two
absorption systems we analyse above. They obtained new \h121cm spectra of
each absorber and compared the measured redshift with the published
molecular redshifts. C00 find $\dy = (+1.0\pm 0.3) \times 10^{-5}$ for
\txs~and $\dy = (+1.29\pm 0.08) \times 10^{-5}$ for \pks~where the errors
are statistical only. They add to this an additional error of $\pm 1\times
10^{-5}$, which is derived from other measurement uncertainties in the
H{\sc \,i} data (e.g. frequency calibration uncertainties). However,
C00 argue that the line of sight problem above could lead to errors as
large as $10{\rm \,kms}^{-1}$ on the basis of typical sub-kiloparsec ISM
motions. They therefore conclude with an upper limit of
$\left|\dot{\alpha}/\alpha\right| < 3.5 \times 10^{-15}{\rm \,yr}^{-1}$
(assuming constant $g_p$) for a look-back time (within their assumed
cosmology) of 4.8\,Gyr to $z=0.6847$. This corresponds to $\left|\dy\right|
< 1.7\times 10^{-5}$. This final dominant error term of $10{\rm
\,kms}^{-1}$ is much greater than the $1.2{\rm \,kms}^{-1}$ error we apply
in equations \ref{eq:resy} and \ref{eq:resa} but we stress that our value
has been obtained empirically and so should be reliable.

As mentioned in Section \ref{sec:intro}, tight constraints on $\da$ come
from optical absorption line studies. One method is to compare the spacing
of single alkali doublets (ADs) in QSO and laboratory spectra (Cowie \&
Songaila 1995; Varshalovich, Panchuk \& Ivanchik 1996). The most precise
constraint to date using this method was obtained by Murphy et al. (2001c),
$\da = (-0.5\pm 1.3)\times 10^{-5}$, where we analysed 21 Si{\sc \,iv}
doublets ($2 < z < 3$) in 13 QSO spectra (see Fig. 3). However, this method
does not take advantage of the large relativistic shifts in the ground state
energy since it compares transitions from the same ground state.

Recently, a new many-multiplet (MM) method was suggested (Dzuba et
al. 1999a,b; Webb et al. 1999) which allows an order of magnitude gain in
precision over the AD method. The magnitude of the relativistic shifts is
sensitive to the nuclear charge and so comparison of transitions in light
and heavy ions yields tight constraints on $\da$.  Also, since any
transition can be used, comparison of transitions with different ground
states is possible. A further increase in precision comes from increased
statistics since one is not restricted to using just a single AD.

The first application of the MM method (Webb et al. 1999) used 30 Mg{\sc
\,i}/Mg{\sc \,ii}/Fe{\sc \,ii} absorption systems ($0.5 < z < 1.6$) towards
17 QSOs and suggested that $\alpha$ may have been smaller in the past: $\da
= (-1.09\pm 0.36) \times 10^{-5}$. More recently, we have re-analysed and
confirmed the Webb et al. results with improved techniques and extended our
analysis to a higher redshift sample of damped Lyman-$\alpha$ systems
(Murphy et al. 2001a; Webb et al. 2001). We compare our new mm/H{\sc \,i}
results with these optical results in Fig. 3 and we see that they are
consistent. The weighted mean for the optical results, $\da = (-0.72 \pm
0.18) \times 10^{-5}$, is significant at the 4.1$\sigma$ level and so a
thorough search for systematic errors has been carried out (Murphy et
al. 2001b). However, no effect was found that could explain the
results. Fig. 3 clearly shows that, assuming a constant $g_p$, a large
sample of mm/H{\sc \,i} absorption systems, over a somewhat larger redshift
range, will allow us to confirm or rule out the change in $\alpha$ implied
by the optical data.

\begin{figure}
\centerline{\psfig{file=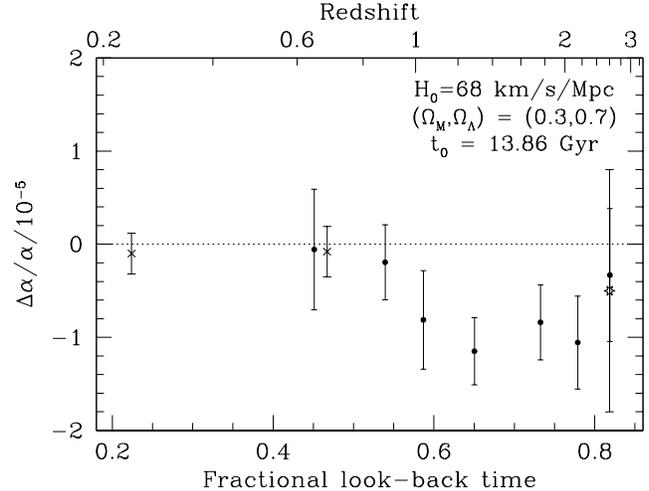,width=3.3in}}
\caption{Comparison of our new mm/H{\sc \,i} constraints (crosses, from
equation \ref{eq:resa}) and our recent optical constraints. The star
represents the weighted mean of $\da$ using 21 Si{\sc \,iv} alkali doublets
from Murphy et al. (2000c) while the dots represent a $7\times 7$ binning
of the 49 absorption systems analysed with the many-multiplet method in
Murphy et al. (2000a).}
\end{figure}

\section*{Acknowledgments}
It is a pleasure to thank Chris Carilli for kindly providing his \h121cm
data for our analysis. We are grateful to the John Templeton Foundation for
supporting this work. MTM and JKW are grateful for hospitality at the IoA
Cambridge, where some of this work was carried out.

\label{lastpage}
\end{document}